# The present status of non-compact lattice QED*

M. Göckeler[a] R. Horsley[a,b] P.E.L. Rakow[c] G. Schierholz[b,a] and H. Stüben[c]

[a]Höchstleistungsrechenzentrum HLRZ, c/o Forschungszentrum Jülich, D-52425 Jülich, Germany

[b]Deutsches Elektronen-Synchrotron DESY, Notkestraße 85, D-22603 Hamburg, Germany

[c]Institut für Theoretische Physik, Freie Universität Berlin, Arnimallee 14, D-14195 Berlin, Germany

We give a 1993 update of non-compact lattice QED, in particular the chiral condensate, finite size effects and meson mass ratios. We compare descriptions of the phase transition. Our previous conclusions remain valid.

## 1. INTRODUCTION

Non-compact lattice QED has become a controversial subject. All are agreed that the theory has a chiral phase transition, but there seems to be little other common ground. Questions that have been raised [1–7] include:
- Where is the phase transition and what are its critical exponents?
- Is the theory interacting or non-interacting at the critical point?
- Measurements of $\alpha_R$ show it decreasing as the phase transition is approached. Is this a sign of triviality, or is the whole charged sector irrelevant due to confinement?
- Are monopoles vital for an understanding of non-compact QED or irrelevant?

In the following sections we shall compare our results and interpretation with other groups.

## 2. CRITICAL BEHAVIOUR OF THE CHIRAL CONDENSATE

In [1] we made a fit to the chiral condensate data available to us in 1991. We used an equation of state that describes the critical region with logarithmically corrected mean field behaviour, the behaviour conventionally expected near phase transitions in $d = 4$ (see also [2])

$$m = \tau \frac{\sigma}{\ln^p(1/\sigma)} + \theta \frac{\sigma^3}{\ln(1/\sigma)}, \qquad (1)$$

where $\sigma$ is the chiral condensate,

$\tau/\theta = \tau_1(1 - \beta/\beta_c)$, $1/\theta = \theta_0 + \theta_1(1 - \beta/\beta_c)$.

*Poster at LAT93, Dallas, U.S.A.

From the fit we found the parameters $\beta_c = 0.186(1)$, $p = 0.61(2)$, $\tau_1 = -0.84(1)$, $\theta_0 = 0.59(1)$ and $\theta_1 = -0.30(2)$.

The equation of state presented in [3] was

$$m = A_\delta \sigma^\delta - A_1(\beta_c - \beta)\sigma, \qquad (2)$$

with the fit parameters $\beta_c = 0.205$, $\delta = 2.31$, $A_\delta = 1.15$ and $A_1 = 5.3125$. The difference between this $\delta$ value and the mean-field value 3 is taken to be evidence of a non-trivial theory.

Since the original fits there is now more data. In Fig. 1 we compare the fit from eq. (1) with an updated set of $\sigma$ measurements. At each point we have shown the data from the largest available lattice. The combined data from [1,3] are shown along with some new measurements. All the $10^4$ and $16^4$ data are new since we made the fit, but we can see that they seem to agree with our expectations. In Fig. 2 we show the same data compared with the power law fit of eq. (2). The fit at lower $m$ values suggests that the true value of $\beta_c$ is lower than the value 0.205.

Important in judging the quality of these fits is an estimate of the finite size effects. Although in general they are small, results in the symmetric phase and from smaller bare masses can suffer from finite size effects. We have modelled the finite size effects by considering the $\beta = \infty$ limit, where they are calculable. We find

$$\sigma - \sigma(\infty) \propto N_f m_R^3 \exp(-2m_R L) + \ldots$$
$$\propto N_f \sigma^3 \exp(-3.23\sigma L) + \ldots \qquad (3)$$

We have used the observation [1] that $\sigma \approx 0.62 m_R$, both at $\beta = \infty$ and in the critical region,

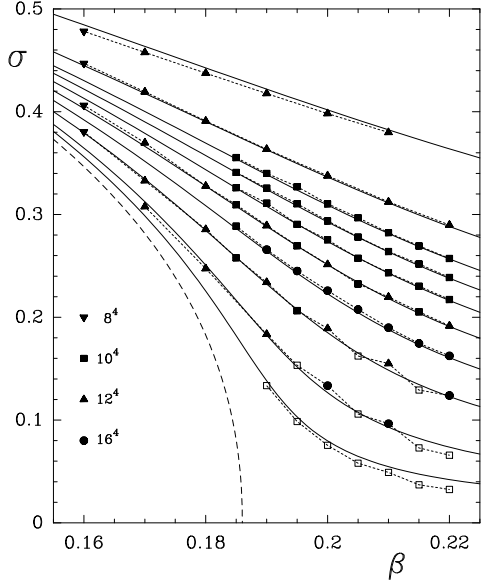

Figure 1. The chiral condensate data and the logarithmically corrected mean field fit. Solid symbols have small finite size effects.

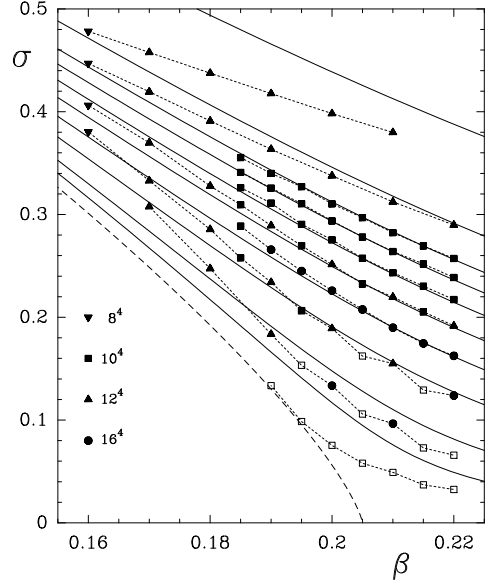

Figure 2. The power law fit and the data.

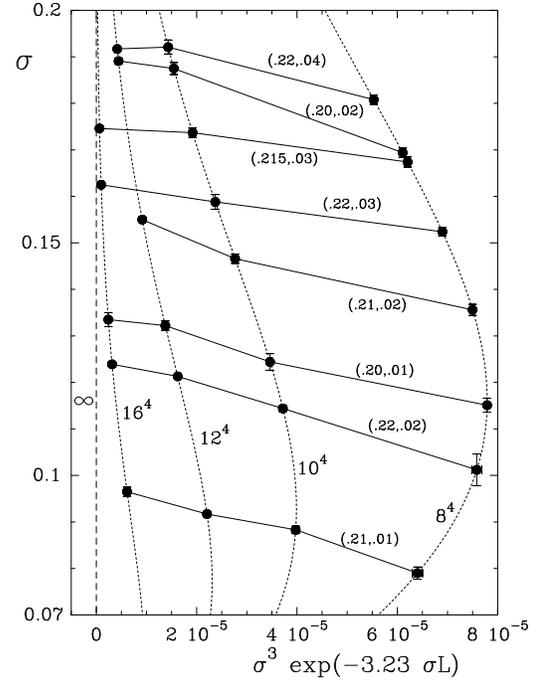

Figure 3. Finite size effects for $\sigma$, shown for various values of $\beta$ and $m$.

to write the formula in terms of $\sigma$, ($m_R$ is the renormalised fermion mass). We now plot $\sigma$ against the right hand side of eq. (3) in Fig. 3. We see reasonably straight lines as predicted. Another observation consistent with eq. (3) is that finite size effects are tiny in the quenched case and grow as the number of flavours, $N_f$, increases [3]. These results allow us to decide when a data point has negligible finite size effects. We estimate that the 'reliable' points are those with $\sigma > 0.3$ for $8^4$ lattices, $\sigma > 0.2$ for $10^4$ lattices and $\sigma > 0.15$ for $12^4$ lattices. For these points finite size effects should be comparable with statistical errors and smaller than the symbols in Figs. (1,2). These measurements are shown as solid symbols.

## 3. MESON MASS RATIOS AND THE CHIRAL CONDENSATE

It has recently been proposed [3,4] that the ratio of the scalar and pseudo-scalar meson masses can be used to test equations of state for the chiral condensate. The argument [4] relating $\partial \ln \sigma / \partial \ln m$ (which we shall call $R_0$) to the meson propagators $C(p)$ and so to the mass ratio

$R_5 \equiv m_\pi^2/m_\sigma^2$ is sketched as:

$$\frac{\partial \ln \sigma}{\partial \ln m} \equiv R_0$$

Ward identities

$$= \frac{C_\sigma(p=0)_{full}}{C_\pi(p=0)_{full}} \equiv R_1$$

Neglect annihilation

$$\stackrel{?}{\approx} \frac{C_\sigma(p=0)_{connect}}{C_\pi(p=0)_{connect}} \equiv R_2$$

One state dominance

$$\stackrel{?}{\approx} \frac{Z_\sigma}{Z_\pi} \frac{(\cosh m_\pi - 1)}{(\cosh m_\sigma - 1)} \equiv R_3$$

Assume $Z_\sigma \approx Z_\pi$

$$\stackrel{?}{\approx} \frac{\cosh m_\pi - 1}{\cosh m_\sigma - 1} \equiv R_4$$

$m_\pi$, $m_\sigma$ small

$$\approx \frac{m_\pi^2}{m_\sigma^2} \equiv R_5$$

We can check most of the steps in this argument, because we can directly measure $R_0$ and all the intermediate quantities except $R_1$ (for which we would need the full meson propagators). Only $R_0$ and $R_1$ can be used as precision tests of an equation of state. $R_2, \ldots, R_5$ are only approximately equal to the derivative of $\sigma$, so the true equation of state *should* deviate from measurements of these quantities. In Fig. 4 we show $R_2$ together with the fit, eq. (1). It deviates somewhat from the measurements of $R_0$, showing that annihilation is not negligible. In Fig. 5 we show the power law fit. $R_2$ does not favour eq. (2). Pictures of $R_5$ can be seen in [3]. It is very different from $R_0$ and $R_2$, agreeing with neither equation of state. Why is $R_5$ so far from $R_2$? The main reasons are the assumption $Z_\pi = Z_\sigma$ and the step from $R_4 \to R_5$. In Fig. 6 we show measurements of both $Z$'s. Away from the critical point the difference can be considerable. In [3,4] it is argued that $R_5$ is a lower bound on $R_0$ on the grounds that $Z_\sigma > Z_\pi$. The data does not agree with this argument.

## 4. THE CHARGED SECTOR

In [1] we looked at the charged sector in some detail, the natural thing to do in QED. In particular we looked at the renormalised charge and found that it decreases as we approach the critical point. (Measuring the strength of physical interactions is a more satisfactory way of discussing triviality than looking at critical exponents.) In [5] it was argued that even non-compact QED

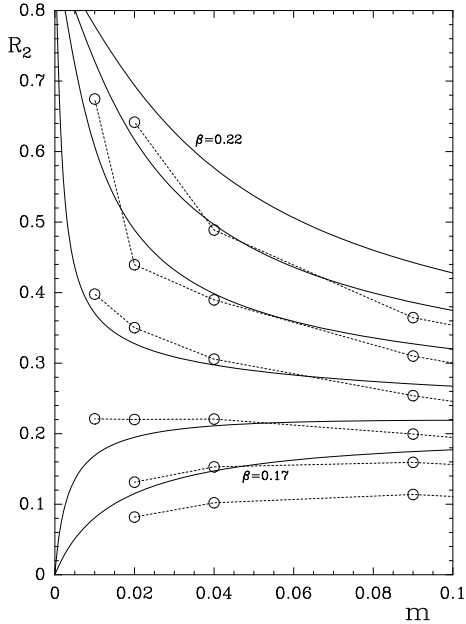

Figure 4. $R_2$ from the propagators together with the logarithmically corrected mean field fit.

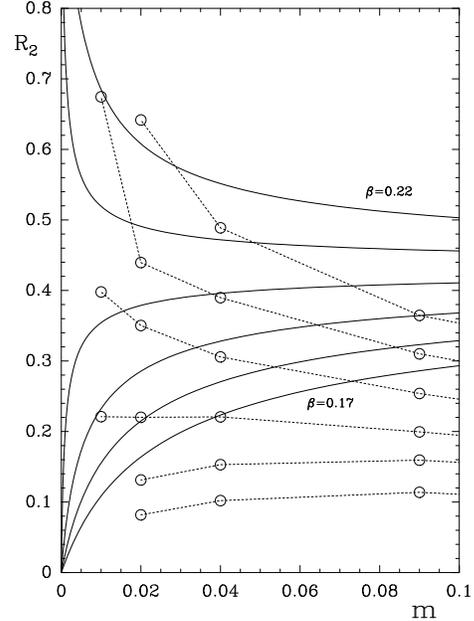

Figure 5. $R_2$ and the power law fit.

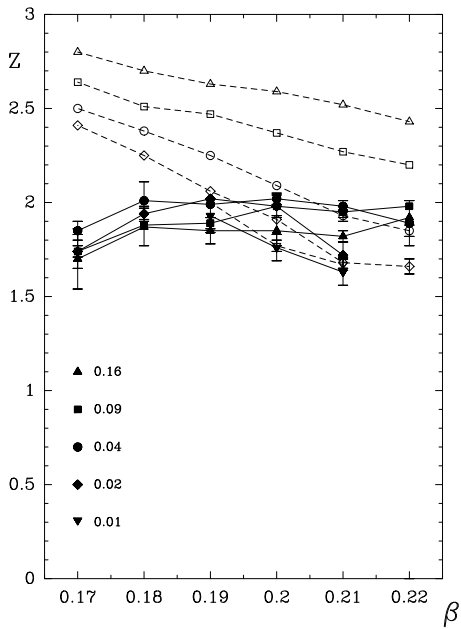

Figure 6. $Z_\pi$ (open symbols) and $Z_\sigma$ (filled symbols) against $\beta$ for different bare masses.

confines due to monopole condensation:

| Cluster Susceptibility | Diverges $\beta \approx 0.205$ |
| --- | --- |
| $\Downarrow$ | $\checkmark$ |
| Monopole Percolation | Occurs $\beta < 0.205$ |
| $\Downarrow$ | ?? |
| Monopole Condensation | Not seen |
| $\Downarrow$ | $\checkmark$ |
| Confinement | Not seen |

The argument only works if all the links hold. The justification of the second step, the connection between the percolation of monopole lines and monopole condensation, is less clear.

We have checked the argument by applying tests for confinement [1] (by looking at the potential and finding a Coulomb behaviour) and for monopole condensation [7] (by using conventional monopole order parameter definitions [8]). Neither is seen. Claims that monopoles are involved in the phase transition are based on the 'exact' coincidence between the monopole percolation threshold and the chiral phase transition. It is hard to make the data for $\sigma$ consistent with this hypothesis (see Figs. (1,2)).

## 5. CONCLUSIONS

In [1] we presented evidence for the triviality of QED. This conclusion was based mainly on measurements of $\alpha_R$, though we also noted that the chiral condensate data are consistent with this hypothesis. This was questioned [3] on the basis of the equation of state in eq. (2), which is plotted in Fig. 2.

It has been argued that the chiral condensate data have severe finite size problems, but comparing data from different lattices we see no large finite size effects. Meson mass ratios have been used to test equations of state [3,4], but approximations must be made to relate mass ratios to the equation of state. We check these approximations against the data, and see that some of them are poor.

## 6. ACKNOWLEDGEMENTS

This work was supported in part by the DFG. The numerical computations were performed on the HLRZ Cray Y-MP in Jülich and the Fujitsu VP 2400 at the RRZN Hannover. We wish to thank these institutions for their support.